\documentclass[aps,prc,twocolumn,superscriptaddress]{revtex4-2}
\usepackage{amsmath}
\usepackage{amssymb}
\usepackage{graphicx}
\usepackage{hyperref}
\hypersetup{
    colorlinks=true,
    linkcolor=blue,
    citecolor=blue,
    urlcolor=blue,
    filecolor=blue
}

\begin{document}

\title{Linear scaling of lepton charge asymmetry in \texorpdfstring{$W^\pm$}~
production in ultra-relativistic nuclear collisions}

\author{An-Ke Lei}
\affiliation{Key Laboratory of Quark and Lepton Physics (MOE) and Institute of
Particle Physics, Central China Normal University, Wuhan, 430079, China}

\author{Dai-Mei Zhou}
\email{zhoudm@mail.ccnu.edu.cn}
\affiliation{Key Laboratory of Quark and Lepton Physics (MOE) and Institute of
Particle Physics, Central China Normal University, Wuhan, 430079, China}

\author{Yu-Liang Yan}
\email{yanyl@ciae.ac.cn}
\affiliation{China Institute of Atomic Energy, P. O. Box 275 (10), Beijing, 102413, China}

\author{Xiao-Ming Zhang}
\affiliation{Key Laboratory of Quark and Lepton Physics (MOE) and Institute of
Particle Physics, Central China Normal University, Wuhan, 430079, China}

\author{Liang Zheng}
\affiliation{School of Mathematics and Physics, China University of Geosciences (Wuhan), Wuhan 430074, China}

\author{Du-Juan Wang}
\affiliation{Department of Physics, Wuhan University of Technology, Wuhan
430070, China}

\author{Xiao-Mei Li}
\affiliation{China Institute of Atomic Energy, P. O. Box 275 (10), Beijing, 102413, China}

\author{Gang Chen}
\affiliation{School of Mathematics and Physics, China University of Geosciences (Wuhan), Wuhan 430074, China}

\author{Xu Cai}
\affiliation{Key Laboratory of Quark and Lepton Physics (MOE) and Institute of
Particle Physics, Central China Normal University, Wuhan, 430079, China}

\author{Ben-Hao Sa}
\email{sabh@ciae.ac.cn}
\affiliation{Key Laboratory of Quark and Lepton Physics (MOE) and Institute of
Particle Physics, Central China Normal University, Wuhan, 430079, China}
\affiliation{China Institute of Atomic Energy, P. O. Box 275 (10), Beijing, 102413, China}

\date{\today}

\begin{abstract}
    The lepton charge asymmetry in $W^\pm$ production in the nuclear collisions at 
$\sqrt{s_{\rm NN}}=5.02$~TeV is investigated with a parton and hadron cascade 
model PACIAE. Recently published ALICE and the ATLAS data of lepton charge 
asymmetry are well reproduced. An interesting linear scaling behavior is 
observed in the lepton charge asymmetry as a function of the collision system 
valence quark number asymmetry among the different size of nuclear collision 
systems at $\sqrt{s_{\rm NN}}=5.02$~TeV. This linear scaling behavior may serve 
as an additional constraint on the PDF (nPDF) extractions.
\end{abstract}

\maketitle

\section{\label{sec:intr} Introduction}

    $W^{\pm}$ vector bosons are heavy particles with masses of $ m_{W^\pm} 
\approx 80$~GeV/$c^{2}$~\cite{ParticleDataGroup:2020ssz,CDF:2022hxs}. They are 
mainly produced in the hard partonic scattering processes with large momentum 
transfer at the early stage of the (ultra-)relativistic nuclear 
collisions. In comparison with evolution time of the heavy-ion collision 
system, 10 to 100~fm/$c$ for instance, the $W^{\pm}$ decay time 
($\sim 0.0922~{\rm fm/}c$, estimated with full decay width, $t=h/\Gamma$ 
\cite{ParticleDataGroup:2020ssz}) is very short. The $W^{\pm}$ leptonic decays 
\begin{equation*}
    W^{\pm}\rightarrow l^{\pm}\nu_{l},~(l{\rm :}~e,\mu,\tau)
\end{equation*}
are nearly instantaneous. As the produced leptons weakly interact with the 
partonic and hadronic matter created in nuclear collisions, the $W^{\pm}$ is a 
powerful probe for investigating the properties of the initial state in the 
nuclear collision system. The main production processes of $W^\pm$ are 
\begin{equation*}
    u\bar{d} \rightarrow W^{+}, \hspace{0.4cm} d\bar{u} \rightarrow W^{-}
\end{equation*}
in the leading-order (LO) approximation~\cite{ParticleDataGroup:2020ssz,Martin:1999ww}.
With the involvement of the valence $u$- and $d$-quarks inside the nuclear 
system, the isospin difference between proton and neutron is important for the 
production of $W^{+}$ and $W^{-}$. It is therefore expected that the $W^\pm$ 
boson charge asymmetry and the corresponding decayed lepton charge asymmetry 
are related to the relative abundance of protons and neutrons (to the relative 
abundance of valence $u$- and $d$-quarks) of the collision system. The lepton 
charge asymmetry $A_{l}$ could be defined as the difference between the 
multiplicity of positive and negative leptons divided by their sum:
\begin{equation}
    A_{l} = \frac{N_{l^{+} \leftarrow W^{+}} - N_{l^{-} \leftarrow W^{-}}}
	    {N_{l^{+} \leftarrow W^{+}} + N_{l^{-} \leftarrow W^{-}}}.
\label{eq:A_l}
\end{equation}
Similarly, the collision system valence $u$ ($d$) quark number asymmetry 
$A_v$ is defined as
\begin{equation}
A_v = \frac{N_u - N_d}{N_u + N_d},
\end{equation}
where $N_u$ ($N_d$) refers to the number of valence quark $u$ ($d$) in 
the nuclear collision system.

    The lepton charge asymmetry have been measured in $p\bar p$ collisions at 
$\sqrt{s}=1.8$ and $1.96$~TeV by CDF~\cite{CDF:1991igh,CDF:1994cas,CDF:1998uzn,CDF:2005cgc,CDF:2009cjw} 
and D0~\cite{D0:2007pcy,D0:2008cgv,D0:2013xqc} collaborations on Fermilab 
Tevatron. They give early complementary constrains on Parton Distribution 
Function (PDF) relative to the deep inelastic scattering experiments. 
More precise measurements in $pp$ collisions at $\sqrt{s}$=7 and 8~TeV are 
reported by ATLAS~\cite{ATLAS:2011pph,ATLAS:2011qdp,ATLAS:2016nqi,ATLAS:2019fgb}, 
CMS~\cite{CMS:2011aa,CMS:2012ivw,CMS:2012fgk,CMS:2013pzl,CMS:2016qqr} and 
LHCb~\cite{LHCb:2012lki,LHCb:2014liz,LHCb:2015mad,LHCb:2016zpq} collaborations 
on the Large Hadron Collider (LHC) at CERN. ATLAS also give the measurements 
in $pp$ collisions at $\sqrt{s}=2.76$~\cite{ATLAS:2019fyu} and $5.02$~TeV
~\cite{ATLAS:2018pyl}. Beyond the elementary $pp$ collisions, the lepton charge 
asymmetry has been measured in the p+Pb collisions at 
$\sqrt{s_{\rm NN}}=5.02$~TeV by ALICE~\cite{ALICE:2016rzo} and at 
$\sqrt{s_{\rm NN}}=8.16$~TeV by ALICE~\cite{ALICE:2022cxs} and CMS~\cite{CMS:2019leu}. 
The results of lepton charge asymmetry in Pb+Pb collisions at 
$\sqrt{s_{\rm NN}}=2.76$~TeV are first given by ATLAS~\cite{ATLAS:2014sic} and 
CMS~\cite{CMS:2012fgk}. Recently, ATLAS and ALICE published their results 
of $W^\pm$ production cross-section and lepton charge asymmetry in Pb+Pb 
collisions at $\sqrt{s_{\rm NN}}=5.02$~TeV~\cite{ATLAS:2019ibd,ALICE:2022cxs}. 
All those measurements are declared to be well reproduced by the LO and/or 
next-to-leading-order (NLO) perturbative Quantum ChromoDynamics (pQCD) 
calculations~\cite{Dulat:2015mca,Eskola:2009uj,Kusina:2016fxy,Eskola:2016oht}
using the CT14 PDF set~\cite{Dulat:2015mca} with or without the nuclear 
modified PDF (nPDF).

    In this paper, the parton and hadron cascade model PACIAE \cite{Sa:2011ye} is 
employed to reproduce the ATLAS and ALICE lepton charge asymmetry data 
in Pb+Pb at $\sqrt{s_{\rm NN}}=5.02$~TeV~\cite{ATLAS:2019ibd,ALICE:2022cxs}. We 
then calculate the lepton charge asymmetry as a function of the collision 
system valence quark number asymmetry in both the elementary nucleon-nucleon 
(NN) and the nucleus-nucleus (AA) collisions at $\sqrt{s_{\rm NN}}=5.02$~TeV. An 
interesting linear scaling behavior in the lepton charge asymmetry as the 
function of collision system valence quark number asymmetry is observed 
among the different NN and AA collision systems.

    The paper is organized as follows. In Sec.~\ref{sec:mod}, we describe the 
PACIAE model and the setups for $W^\pm$ production in \textsc{Pythia} 6.4, 
with a precision equivalent to the NLO pQCD. The simulated results and 
discussions are presented in Sec.~\ref{sec:res}. At last, a summary is 
given in Sec.~\ref{sec:summ}.

\section{\label{sec:mod} Model}

    A parton and hadron cascade model PACIAE~\cite{Sa:2011ye} is employed to 
simulate the elementary nucleon-nucleon (NN: $pp$, $pn$, $np$ and $nn$) 
and the nucleus-nucleus (AA: Cu+Cu, Xe+Xe, Au+Au, Pb+Pb and U+U) 
collisions at $\sqrt{s_{\rm NN}}=5.02$~TeV.

    The PACIAE model is based on \textsc{Pythia} event generator (version $6.4.28$) 
\cite{Sjostrand:2006za}. For NN collisions, with respect to \textsc{Pythia}, 
the partonic and hadronic rescatterings are introduced before and after 
the hadronization, respectively. The final hadronic state is developed 
from the initial partonic hard scattering and parton showers, followed by parton
rescattering, string fragmentation, and hadron rescattering stages. Thus, 
the PACIAE model provides a multi-stage transport description on the evolution 
of the NN collision system.

    For AA collisions, the initial positions of nucleons in the colliding nucleus 
are sampled according to the Woods-Saxon distribution. Together with the 
initial momentum setup of $p_{x} = p_{y} = 0$ and $p_{z} =p_{\rm beam}$ for 
each nucleon, a list containing the initial state of all nucleons in a given 
AA collision is constructed. A collision happened between two nucleons 
from different nuclei if their relative transverse distance is less than or 
equal to the minimum approaching distance:
$D\leq\sqrt{\sigma_{\rm NN}^{\rm tot}/\pi}$. The collision time is calculated 
with the assumption of straight-line trajectories. All such nucleon pairs 
compose an NN collision time list. An NN collision with least collision time is 
selected from the list and executed by \textsc{Pythia} (PYEVNW subroutine) with 
the hadronization temporarily turned-off, as well as the strings and diquarks 
broken-up. The nucleon list and NN collision time list are then updated. A new 
NN collision with least collision time is selected from the updated NN collision 
time list and executed by \textsc{Pythia}. With repeating the aforementioned 
steps till the NN collision list empty, the initial partonic state is 
constructed for a AA collision.

    Then, the partonic rescatterings are performed, where the LO-pQCD parton-parton 
cross section~\cite{Combridge:1977dm,field} is employed. After partonic 
rescattering, the string is recovered and then hadronized with the Lund string 
fragmentation scheme resulting in an intermediate hadronic state. Finally, the 
system proceeds into the hadronic rescattering stage and produces the final 
hadronic state observed in the experiments.

    The PACIAE Monte-Carlo simulation provides a complete description of the NN 
and AA collisions, which includes the partonic initialization, partonic 
rescattering, hadronization, and the hadronic rescattering stages. Meanwhile, 
a PACIAE model simulation could be selected to stop at any stage desired 
conveniently. In this work, the simulations are stopped at the partonic 
initialization stage, after the initial hard scattering and parton shower 
processes happened.

    In this study, the $W^\pm$ relevant production channels are activated in a 
user-controlled approach by setting ${\rm MSEL}=0$ in $\textsc{Pythia}$, 
together with the following subprocesses switched-on:
\begin{align*}
    f_{i}\overline{f}_{j} & \rightarrow W^{+}/W^{-},   \\
    f_{i}\overline{f}_{j} & \rightarrow g W^{+}/W^{-}, \\
    f_{i}\overline{f}_{j} & \rightarrow \gamma W^{+}/W^{-}.
\end{align*}
In the above equations, $f$ refers to fermions (quarks) and its subscript 
stands for the flavor code.

    In addition, the parton shower is switched off by setting MSTP(61)=0 and 
MSTP(71)=0, to prevent the double-counting. 
The PDF and other parameters are kept the same as the default \textsc{Pythia}. 
Thus the PACIAE (\textsc{Pythia}) model simulation is corresponding to the NLO 
pQCD in precision.

\section{\label{sec:res} Results and Discussions}

\begin{figure*}[htbp]
    \includegraphics[width=0.8\textwidth]{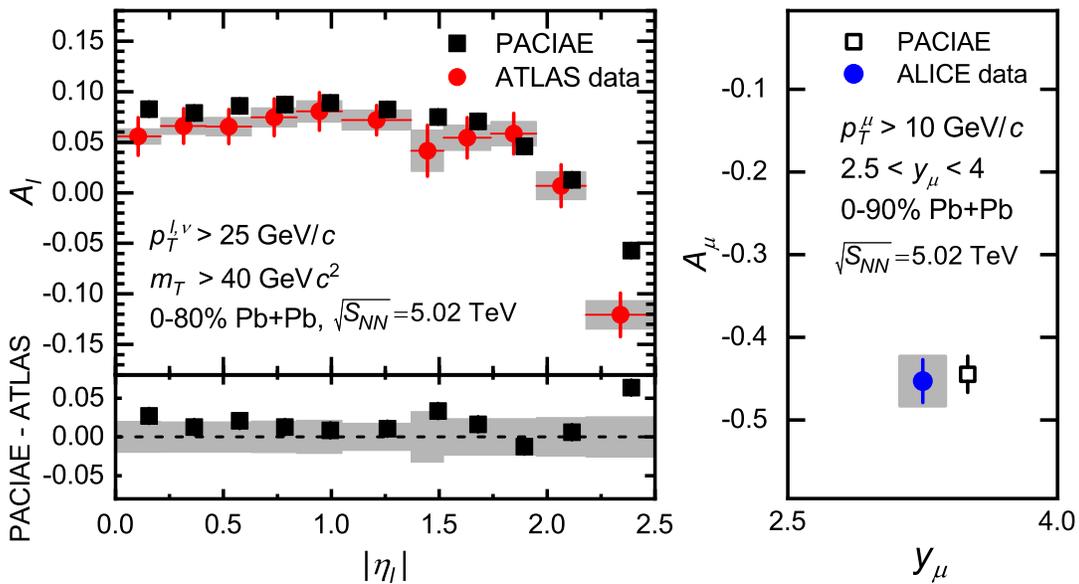}
    \caption{\label{fig:comp} Left panel: the combined electron and muon 
charge asymmetry as a function of absolute pseudorapidity for the 0-80\% 
centrality class in Pb+Pb collisions at $\sqrt{s_{\rm NN}}=5.02$~TeV. The black 
solid squares with vertical error bar (hard to see) show the results of PACIAE 
simulations. The red solid circles with vertical error bars and shaded error 
boxes represent the ATLAS data ~\cite{ATLAS:2019ibd}. The lower sub-panel shows 
the differences between the PACIAE calculations and the ATLAS data, with the 
shaded boxes representing the total experimental uncertainties. The 
PACIAE result points are shifted horizontally for better visibility.
Right panel: the muon charge asymmetry for the 0-90\% centrality class in 
Pb+Pb collisions at $\sqrt{s_{\rm NN}}=5.02$~TeV, with kinematic cuts of 
$p^{\mu}_T > 10$~GeV/$c$ and $2.5 < y_{\mu} < 4.0$. Here the open square with 
vertical error bar shows the result of PACIAE simulation, and the blue solid 
circle with vertical error bar and shaded error box represents ALICE 
darum~\cite{ALICE:2022cxs}.}
\end{figure*}

    In the left panel of Fig.~\ref{fig:comp}, we compare the PACIAE model 
calculated electron and muon combined lepton charge asymmetry (black solid 
squares) with ATLAS data~\cite{ATLAS:2019ibd} (red solid circles) for the 
0-80\% centrality class in Pb+Pb collisions at $\sqrt{s_{\rm NN}}=5.02$~TeV. 
The bottom sub-panel shows the differences between PACIAE calculation and ATLAS 
data, where the shaded boxes show the total uncertainties in experimental data. 
The kinematic cuts in PACIAE simulation is the same as the ones in ATLAS 
experiments~\cite{ATLAS:2019ibd}, i.e.
\begin{align*}
    p^{l}_T > 25 {\rm~GeV}/c, & \hspace{0.4cm} p^{\nu}_T > 25 {\rm~GeV}/c, \\
    |\eta_l| < 2.5, & \hspace{0.4cm} m_T > 40 {\rm~GeV}/c^2,
\end{align*}
where $p^{l}_T$ and $p^{\nu}_T$ stand for the transverse momentum of the 
lepton and the corresponding neutrino, respectively. 

\noindent $|\eta_l|$ refers to the
absolute pseudorapidity of the lepton, and $m_T$ is the transverse mass of the 
lepton and neutrino system. 
The pseudorapidity bins are divided as follows:
\begin{align*}
    |\eta_l| = &~(0,0.21),~~~~~(0.21,0.42), ~(0.42,0.63), ~(0.63,0.84), \\
               &~(0.84,1.05), ~(1.05,1.37), ~(1.37,1.52), ~(1.52,1.74), \\
               &~(1.74,1.95), ~(1.95,2.18), ~(2.18,2.5).
\end{align*}

    We see in this panel, 
the PACIAE calculated lepton charge asymmetry is well consistent with ATLAS 
data~\cite{ATLAS:2019ibd}, except the most forward $|\eta_l|$ bin. This 
exception is also existed in the pQCD predictions calculated with the CT14 NLO 
PDF set, even in the EPPS16 and nCTEQ15 nPDF sets~\cite{ATLAS:2019ibd}.

    The PACIAE calculation of the muon charge asymmetry (open square) for the 
0-90\%  centrality class in Pb+Pb collisions at $\sqrt{s_{\rm NN}}=5.02$~TeV is 
compared with the ALICE datum ~\cite{ALICE:2022cxs} (blue solid circle) in the 
right panel of Fig.~\ref{fig:comp}. The same kinematic cuts of 
$p^{\mu}_T > 10$~GeV/$c$ and $2.5 < y_{\mu} < 4$~\footnote[1]
{The ALICE muon spectrometer covers negative pseudorapidity. However, in 
symmetric Pb+Pb collisions, positive values of rapidity are conventionally
used for the muon coverage.} 
are set in both the experiment and theory. One can see in this panel that, the 
calculated result agrees well with the ALICE datum. 

\begin{figure}[htbp]
    \includegraphics[width=0.45\textwidth]{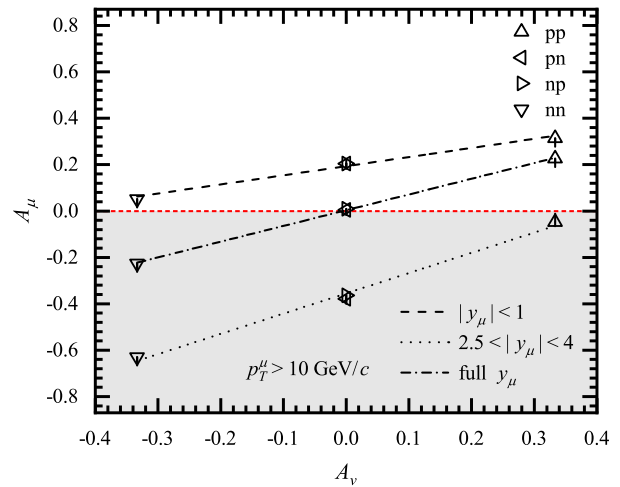}
    \caption{\label{fig:Amu_pp} The muon charge asymmetry as a function of the 
valence quark asymmetry in the elementary NN collisions ($pp$, open 
triangle-ups; $pn$, open triangle-lefts; $np$, open triangle-rights; 
and $nn$, open triangle-downs) at $\sqrt{s}=5.02$~TeV for three rapidity 
intervals of $|y|<1$, $2.5<|y|<4.0$, and full rapidity phase-space. The dashed, 
dotted, and dash-dotted lines are the corresponding linear fitting to the $A_\mu$ 
vs. $A_v$ functions, respectively. The kinematic cut is $p^\mu_T > 10$~GeV/$c$. 
The red short-dashed zero-baseline and the shaded area are added for better 
distinguishment between the positive and negative values of $A_\mu$.}
\end{figure}

    In Fig.~\ref{fig:Amu_pp}, we give the muon charge asymmetry as a function of 
the valence quark number asymmetry in the elementary NN collisions of $pp$ 
(open triangle-ups), $pn$ (triangle-lefts), $np$ (triangle-rights), and $nn$ 
(triangle-downs) 

\noindent at $\sqrt{s}=5.02$~TeV. The kinematic cut of $p^\mu_T > 
10$~GeV/$c$ is implemented in the simulations for three $y_\mu$ intervals: 
$|y_\mu|<1$, $2.5<|y_\mu|<4.0$, and full $y_\mu$ phase-space. 
The dashed, dotted and dash-dotted lines represent the linear fitting to the 
function of $A_\mu$ vs. $A_v$ in three $y_\mu$ intervals above, respectively. 
One can see all of $A_\mu$ in $pp$, $pn$, $np$ and $nn$ collisions are positive 
in mid-rapidity ($|y_\mu| < 1$), while negative in the forward rapidity 
($2.5 < |y_\mu| <4$). This sign-changing phenomenon has been also observed in 
LHCb measurements~\cite{LHCb:2012lki,LHCb:2014liz,LHCb:2015mad,LHCb:2016zpq}. 
It comes from the different helicity dependence of the lepton couplings to the 
boson~\cite{LHCb:2012lki}. For full rapidity phase-space, $A_\mu$ in $pn$ and 
$np$ collisions is consistent with zero owing to the vanishing relative numbers 
between $u$- and $d$-valence quarks, while the $A_\mu$ in $pp$ and $nn$ are 
almost equal in magnitude but opposite in sign. On the other hand, we can see 
the slope fitted in mid-rapidity is gentler than the one in the forward 
rapidity, due to the isospin effect in $W^{\pm}$ production is more pronounced 
at forward rapidity~\cite{ConesaDelValle:2007dza}.

\begin{figure}[htbp]
    \includegraphics[width=0.45\textwidth]{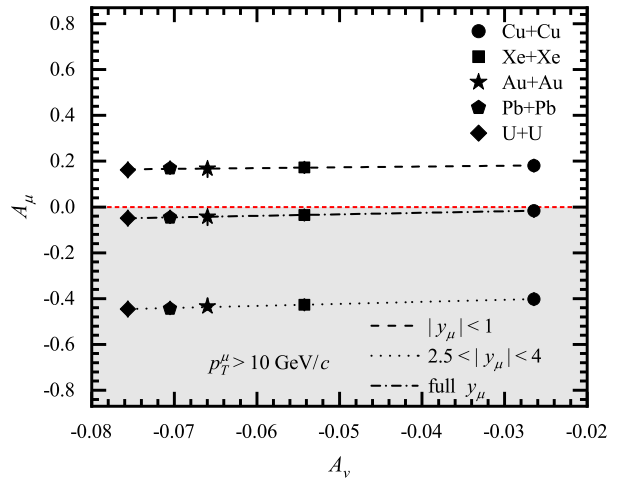}
    \caption{\label{fig:Amu_AA} The muon charge asymmetry as a function of 
the valence quark number asymmetry in the symmetric AA collisions (Cu+Cu, solid 
circles; Xe+Xe, solid squares; Au+Au, solid stars; Pb+Pb, solid pentagons; and 
U+U, solid diamonds) at $\sqrt{s_{\rm NN}}=5.02$~TeV for three rapidity 
intervals of $|y_\mu|<1$, $2.5<|y_\mu|<4.0$, and full rapidity phase-space. The 
dashed, dotted and dash-dotted lines are the corresponding linear fitting to the 
$A_\mu$ vs. $A_v$ functions, respectively. The kinematic cut is 
$p^\mu_T > 10$~GeV/$c$. The red short-dashed zero-baseline and the shaded area 
are added for better distinguishment between the positive and negative values 
of $A_\mu$.}

\end{figure}
\begin{figure}[htbp]
    \includegraphics[width=0.45\textwidth]{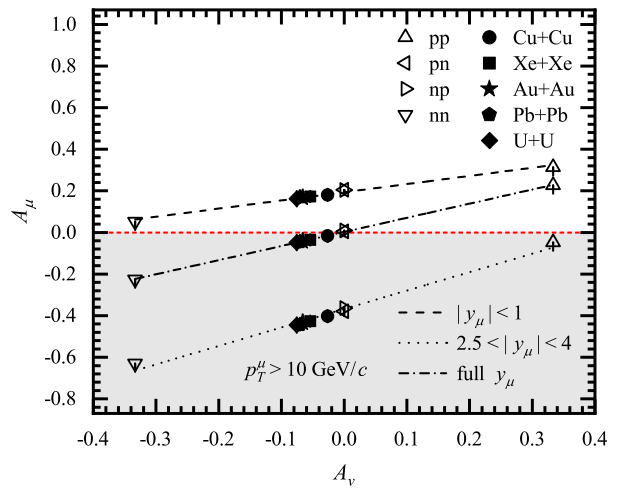}
    \caption{\label{fig:Amu_AApp} The muon charge asymmetry as a function of 
the valence quark number asymmetry in the elementary NN collisions and the symmetric 
AA collisions at $\sqrt{s_{\rm NN}}=5.02$~TeV, obtained by putting 
Fig.~\ref{fig:Amu_pp} and \ref{fig:Amu_AA} together.}
\end{figure}

    The correlation between $A_\mu$ and $A_v$ in symmetrical AA collisions of Cu+Cu 
(solid circles), Xe+Xe (solid squares), Au+Au (solid stars), Pb+Pb (solid 
pentagons) and U+U (solid diamonds) at $\sqrt{s_{\rm NN}}=5.02$~TeV for three 
rapidity intervals is shown in Fig.~\ref{fig:Amu_AA}, with the kinematic cut of 
$p^\mu_T > 10$~GeV/$c$. The corresponding linear fitting is also given. 
In this figure, the valence quark number asymmetry ($A_v$) are all negative, 
since the number of neutrons (the number of $d$ quarks) is always greater than 
that of protons ($u$ quarks) in heavy nuclei. The $A_\mu$ is positive 
in the mid-rapidity bin for all five AA collision systems but negative 
in the forward and full rapidity regions. It can be attributed to the $V-A$ 
decays of $W$ boson. This structure also affects the slope of the linear 
fitting. With the ratio of neutron to proton ($A_v$ magnitude) increasing, the 
fraction of $W^- \rightarrow l^- \nu_l$ becomes larger in forward rapidity 
resulting a steeper slope than the one in mid-rapidity~\cite{LHCb:2012lki,ConesaDelValle:2007dza}. 
It is more clear in Fig.~\ref{fig:Amu_AApp}, where we draw the $A_\mu$ vs. $A_v$ 
function in all the NN and AA collisions together. Here we see the fitted slope 
in the full rapidity is larger than the one in mid-rapidity, but is smaller than 
that in forward rapidity. One more interesting feature is that, the linear 
scaling behavior of $A_\mu$ vs. $A_v$ shown in Fig.~\ref{fig:Amu_pp} is also 
observed in Fig.~\ref{fig:Amu_AA} and~\ref{fig:Amu_AApp}.

\section{\label{sec:summ} Summary}

    In summary, we studied the lepton charge asymmetry from the $W^\pm$ decays in 
both the NN and AA nuclear collisions at $\sqrt{s_{\rm NN}}=5.02$~TeV in frame 
of the Monte-Carlo simulation model PACIAE. The simulated results well 
reproduce the corresponding ATLAS and ALICE data. Furthermore, the correlation 
between the lepton charge asymmetry $A_\mu$ and the valence quark number 
asymmetry $A_v$ is studied in the above nuclear collisions for three rapidity 
intervals: $|y_\mu|<1$, $2.5 < |y_\mu| < 4$, and the full rapidity phase space. 
In each rapidity interval, a linear scaling behavior of $A_\mu$ vs. $A_v$ 
function is observed in the elementary NN collision systems 
($pp$, $pn$, $np$, and $nn$), the AA collision systems 
(Cu+Cu, Xe+Xe, Au+Au, Pb+Pb, and U+U), and even in their combination.

    The universal linear scaling behavior between $A_\mu$ and $A_v$ explored in 
both nucleon-nucleon and nucleus-nucleus collision systems may provide an 
additional constraint in the extraction of PDF (nPDF). On the contrary, we 
also note that the nuclear modifications on the vacuum nucleon PDF change the 
in-going parton kinematics in the hard scattering processes and consequently 
impact the $p_T$- and $y$-distribution of the out-going particles. This effect 
should result in a breaking of the $A_\mu$ vs. $A_v$ scaling established if 
looking into a given lepton $p_T$--$y$ phase space. Therefore, by measuring 
$p_T$ and $y$ differentially, the deviation of the $A_\mu$ vs. $A_v$ 
correlations between NN and AA collision systems should provide a new tool to 
constrain the nPDF in a wide Bjorken-$x$ region. This has to be studied 
further.

\begin{acknowledgments}
    The authors thank helpful discussion with Ming-Rui Zhao. This work was 
supported by the National Natural Science Foundation of China (11775094, 
11805079, 11905188, 11775313, 11905163), the Continuous Basic Scientific 
Research Project (No.WDJC-2019-16) in CIAE, National Key Research and 
Development Project (2018YFE0104800) and by the 111 project of the foreign 
expert bureau of China.
\end{acknowledgments}


\begin{thebibliography}{41}%
\makeatletter
\providecommand \@ifxundefined [1]{%
 \@ifx{#1\undefined}
}%
\providecommand \@ifnum [1]{%
 \ifnum #1\expandafter \@firstoftwo
 \else \expandafter \@secondoftwo
 \fi
}%
\providecommand \@ifx [1]{%
 \ifx #1\expandafter \@firstoftwo
 \else \expandafter \@secondoftwo
 \fi
}%
\providecommand \natexlab [1]{#1}%
\providecommand \enquote  [1]{``#1''}%
\providecommand \bibnamefont  [1]{#1}%
\providecommand \bibfnamefont [1]{#1}%
\providecommand \citenamefont [1]{#1}%
\providecommand \href@noop [0]{\@secondoftwo}%
\providecommand \href [0]{\begingroup \@sanitize@url \@href}%
\providecommand \@href[1]{\@@startlink{#1}\@@href}%
\providecommand \@@href[1]{\endgroup#1\@@endlink}%
\providecommand \@sanitize@url [0]{\catcode `\\12\catcode `\$12\catcode
  `\&12\catcode `\#12\catcode `\^12\catcode `\_12\catcode `\%12\relax}%
\providecommand \@@startlink[1]{}%
\providecommand \@@endlink[0]{}%
\providecommand \url  [0]{\begingroup\@sanitize@url \@url }%
\providecommand \@url [1]{\endgroup\@href {#1}{\urlprefix }}%
\providecommand \urlprefix  [0]{URL }%
\providecommand \Eprint [0]{\href }%
\providecommand \doibase [0]{https://doi.org/}%
\providecommand \selectlanguage [0]{\@gobble}%
\providecommand \bibinfo  [0]{\@secondoftwo}%
\providecommand \bibfield  [0]{\@secondoftwo}%
\providecommand \translation [1]{[#1]}%
\providecommand \BibitemOpen [0]{}%
\providecommand \bibitemStop [0]{}%
\providecommand \bibitemNoStop [0]{.\EOS\space}%
\providecommand \EOS [0]{\spacefactor3000\relax}%
\providecommand \BibitemShut  [1]{\csname bibitem#1\endcsname}%
\let\auto@bib@innerbib\@empty
\bibitem [{\citenamefont {Zyla}\ \emph {et~al.}(2020)\citenamefont {Zyla} \emph
  {et~al.}}]{ParticleDataGroup:2020ssz}%
  \BibitemOpen
  \bibfield  {author} {\bibinfo {author} {\bibfnamefont {P.~A.}\ \bibnamefont
  {Zyla}} \emph {et~al.} (\bibinfo {collaboration} {Particle Data Group}),\
  }\bibfield  {title} {\bibinfo {title} {{Review of Particle Physics}},\ }\href
  {https://doi.org/10.1093/ptep/ptaa104} {\bibfield  {journal} {\bibinfo
  {journal} {PTEP}\ }\textbf {\bibinfo {volume} {2020}},\ \bibinfo {pages}
  {083C01} (\bibinfo {year} {2020})}\BibitemShut {NoStop}%
\bibitem [{\citenamefont {Aaltonen}\ \emph {et~al.}(2022)\citenamefont
  {Aaltonen} \emph {et~al.}}]{CDF:2022hxs}%
  \BibitemOpen
  \bibfield  {author} {\bibinfo {author} {\bibfnamefont {T.}~\bibnamefont
  {Aaltonen}} \emph {et~al.} (\bibinfo {collaboration} {CDF}),\ }\bibfield
  {title} {\bibinfo {title} {{High-precision measurement of the W boson mass
  with the CDF II detector}},\ }\href {https://doi.org/10.1126/science.abk1781}
  {\bibfield  {journal} {\bibinfo  {journal} {Science}\ }\textbf {\bibinfo
  {volume} {376}},\ \bibinfo {pages} {170} (\bibinfo {year}
  {2022})}\BibitemShut {NoStop}%
\bibitem [{\citenamefont {Martin}\ \emph {et~al.}(2000)\citenamefont {Martin},
  \citenamefont {Roberts}, \citenamefont {Stirling},\ and\ \citenamefont
  {Thorne}}]{Martin:1999ww}%
  \BibitemOpen
  \bibfield  {author} {\bibinfo {author} {\bibfnamefont {A.~D.}\ \bibnamefont
  {Martin}}, \bibinfo {author} {\bibfnamefont {R.~G.}\ \bibnamefont {Roberts}},
  \bibinfo {author} {\bibfnamefont {W.~J.}\ \bibnamefont {Stirling}},\ and\
  \bibinfo {author} {\bibfnamefont {R.~S.}\ \bibnamefont {Thorne}},\ }\bibfield
   {title} {\bibinfo {title} {{Parton distributions and the LHC: $W$ and $Z$
  production}},\ }\href {https://doi.org/10.1007/s100520050740} {\bibfield
  {journal} {\bibinfo  {journal} {Eur. Phys. J. C}\ }\textbf {\bibinfo {volume}
  {14}},\ \bibinfo {pages} {133} (\bibinfo {year} {2000})},\ \Eprint
  {https://arxiv.org/abs/hep-ph/9907231} {arXiv:hep-ph/9907231} \BibitemShut
  {NoStop}%
\bibitem [{\citenamefont {Abe}\ \emph {et~al.}(1992)\citenamefont {Abe} \emph
  {et~al.}}]{CDF:1991igh}%
  \BibitemOpen
  \bibfield  {author} {\bibinfo {author} {\bibfnamefont {F.}~\bibnamefont
  {Abe}} \emph {et~al.} (\bibinfo {collaboration} {CDF}),\ }\bibfield  {title}
  {\bibinfo {title} {{Lepton asymmetry in W decays from $\bar{p}p$ collisions
  at $\sqrt{s} = 1.8$ TeV}},\ }\href
  {https://doi.org/10.1103/PhysRevLett.68.1458} {\bibfield  {journal} {\bibinfo
   {journal} {Phys. Rev. Lett.}\ }\textbf {\bibinfo {volume} {68}},\ \bibinfo
  {pages} {1458} (\bibinfo {year} {1992})}\BibitemShut {NoStop}%
\bibitem [{\citenamefont {Abe}\ \emph {et~al.}(1995)\citenamefont {Abe} \emph
  {et~al.}}]{CDF:1994cas}%
  \BibitemOpen
  \bibfield  {author} {\bibinfo {author} {\bibfnamefont {F.}~\bibnamefont
  {Abe}} \emph {et~al.} (\bibinfo {collaboration} {CDF}),\ }\bibfield  {title}
  {\bibinfo {title} {{The Charge asymmetry in $W$ boson decays produced in
  $p\bar{p}$ collisions at $\sqrt{s} = 1.8$ TeV}},\ }\href
  {https://doi.org/10.1103/PhysRevLett.74.850} {\bibfield  {journal} {\bibinfo
  {journal} {Phys. Rev. Lett.}\ }\textbf {\bibinfo {volume} {74}},\ \bibinfo
  {pages} {850} (\bibinfo {year} {1995})},\ \Eprint
  {https://arxiv.org/abs/hep-ex/9501008} {arXiv:hep-ex/9501008} \BibitemShut
  {NoStop}%
\bibitem [{\citenamefont {Abe}\ \emph {et~al.}(1998)\citenamefont {Abe} \emph
  {et~al.}}]{CDF:1998uzn}%
  \BibitemOpen
  \bibfield  {author} {\bibinfo {author} {\bibfnamefont {F.}~\bibnamefont
  {Abe}} \emph {et~al.} (\bibinfo {collaboration} {CDF}),\ }\bibfield  {title}
  {\bibinfo {title} {{Measurement of the Lepton Charge Asymmetry in $W$ Boson
  Decays Produced in $p \bar{p}$ Collisions}},\ }\href
  {https://doi.org/10.1103/PhysRevLett.81.5754} {\bibfield  {journal} {\bibinfo
   {journal} {Phys. Rev. Lett.}\ }\textbf {\bibinfo {volume} {81}},\ \bibinfo
  {pages} {5754} (\bibinfo {year} {1998})},\ \Eprint
  {https://arxiv.org/abs/hep-ex/9809001} {arXiv:hep-ex/9809001} \BibitemShut
  {NoStop}%
\bibitem [{\citenamefont {Acosta}\ \emph {et~al.}(2005)\citenamefont {Acosta}
  \emph {et~al.}}]{CDF:2005cgc}%
  \BibitemOpen
  \bibfield  {author} {\bibinfo {author} {\bibfnamefont {D.}~\bibnamefont
  {Acosta}} \emph {et~al.} (\bibinfo {collaboration} {CDF}),\ }\bibfield
  {title} {\bibinfo {title} {{Measurement of the forward-backward charge
  asymmetry from $W \to e \nu$ production in $p\bar{p}$ collisions at $\sqrt{s}
  = 1.96$ TeV}},\ }\href {https://doi.org/10.1103/PhysRevD.71.051104}
  {\bibfield  {journal} {\bibinfo  {journal} {Phys. Rev. D}\ }\textbf {\bibinfo
  {volume} {71}},\ \bibinfo {pages} {051104} (\bibinfo {year} {2005})},\
  \Eprint {https://arxiv.org/abs/hep-ex/0501023} {arXiv:hep-ex/0501023}
  \BibitemShut {NoStop}%
\bibitem [{\citenamefont {Aaltonen}\ \emph {et~al.}(2009)\citenamefont
  {Aaltonen} \emph {et~al.}}]{CDF:2009cjw}%
  \BibitemOpen
  \bibfield  {author} {\bibinfo {author} {\bibfnamefont {T.}~\bibnamefont
  {Aaltonen}} \emph {et~al.} (\bibinfo {collaboration} {CDF}),\ }\bibfield
  {title} {\bibinfo {title} {{Direct Measurement of the $W$ Production Charge
  Asymmetry in $p\bar{p}$ Collisions at $\sqrt{s} = 1.96$ TeV}},\ }\href
  {https://doi.org/10.1103/PhysRevLett.102.181801} {\bibfield  {journal}
  {\bibinfo  {journal} {Phys. Rev. Lett.}\ }\textbf {\bibinfo {volume} {102}},\
  \bibinfo {pages} {181801} (\bibinfo {year} {2009})},\ \Eprint
  {https://arxiv.org/abs/0901.2169} {arXiv:0901.2169 [hep-ex]} \BibitemShut
  {NoStop}%
\bibitem [{\citenamefont {Abazov}\ \emph
  {et~al.}(2008{\natexlab{a}})\citenamefont {Abazov} \emph
  {et~al.}}]{D0:2007pcy}%
  \BibitemOpen
  \bibfield  {author} {\bibinfo {author} {\bibfnamefont {V.~M.}\ \bibnamefont
  {Abazov}} \emph {et~al.} (\bibinfo {collaboration} {D0}),\ }\bibfield
  {title} {\bibinfo {title} {{Measurement of the muon charge asymmetry from $W$
  boson decays}},\ }\href {https://doi.org/10.1103/PhysRevD.77.011106}
  {\bibfield  {journal} {\bibinfo  {journal} {Phys. Rev. D}\ }\textbf {\bibinfo
  {volume} {77}},\ \bibinfo {pages} {011106} (\bibinfo {year}
  {2008}{\natexlab{a}})},\ \Eprint {https://arxiv.org/abs/0709.4254}
  {arXiv:0709.4254 [hep-ex]} \BibitemShut {NoStop}%
\bibitem [{\citenamefont {Abazov}\ \emph
  {et~al.}(2008{\natexlab{b}})\citenamefont {Abazov} \emph
  {et~al.}}]{D0:2008cgv}%
  \BibitemOpen
  \bibfield  {author} {\bibinfo {author} {\bibfnamefont {V.~M.}\ \bibnamefont
  {Abazov}} \emph {et~al.} (\bibinfo {collaboration} {D0}),\ }\bibfield
  {title} {\bibinfo {title} {{Measurement of the electron charge asymmetry in
  $p \bar{p} \to W + X \to e \nu + X$ events at $\sqrt{s}$ = 1.96-TeV}},\
  }\href {https://doi.org/10.1103/PhysRevLett.101.211801} {\bibfield  {journal}
  {\bibinfo  {journal} {Phys. Rev. Lett.}\ }\textbf {\bibinfo {volume} {101}},\
  \bibinfo {pages} {211801} (\bibinfo {year} {2008}{\natexlab{b}})},\ \Eprint
  {https://arxiv.org/abs/0807.3367} {arXiv:0807.3367 [hep-ex]} \BibitemShut
  {NoStop}%
\bibitem [{\citenamefont {Abazov}\ \emph {et~al.}(2013)\citenamefont {Abazov}
  \emph {et~al.}}]{D0:2013xqc}%
  \BibitemOpen
  \bibfield  {author} {\bibinfo {author} {\bibfnamefont {V.~M.}\ \bibnamefont
  {Abazov}} \emph {et~al.} (\bibinfo {collaboration} {D0}),\ }\bibfield
  {title} {\bibinfo {title} {{Measurement of the Muon Charge Asymmetry in
  $p\bar{p}$ $\to$ W+X $\to$ $\mu\nu$ + X Events at $\sqrt{s}$=1.96 TeV}},\
  }\href {https://doi.org/10.1103/PhysRevD.88.091102} {\bibfield  {journal}
  {\bibinfo  {journal} {Phys. Rev. D}\ }\textbf {\bibinfo {volume} {88}},\
  \bibinfo {pages} {091102} (\bibinfo {year} {2013})},\ \Eprint
  {https://arxiv.org/abs/1309.2591} {arXiv:1309.2591 [hep-ex]} \BibitemShut
  {NoStop}%
\bibitem [{\citenamefont {Aad}\ \emph {et~al.}(2011)\citenamefont {Aad} \emph
  {et~al.}}]{ATLAS:2011pph}%
  \BibitemOpen
  \bibfield  {author} {\bibinfo {author} {\bibfnamefont {G.}~\bibnamefont
  {Aad}} \emph {et~al.} (\bibinfo {collaboration} {ATLAS}),\ }\bibfield
  {title} {\bibinfo {title} {{Measurement of the $W$ charge asymmetry in the $W
  \to \mu \nu$ decay mode in $pp$ collisions at $\sqrt s=7$ TeV with the ATLAS
  detector}},\ }\href {https://doi.org/10.1016/j.physletb.2011.05.024}
  {\bibfield  {journal} {\bibinfo  {journal} {Phys. Lett. B}\ }\textbf
  {\bibinfo {volume} {701}},\ \bibinfo {pages} {31} (\bibinfo {year} {2011})},\
  \Eprint {https://arxiv.org/abs/1103.2929} {arXiv:1103.2929 [hep-ex]}
  \BibitemShut {NoStop}%
\bibitem [{\citenamefont {Aad}\ \emph {et~al.}(2012)\citenamefont {Aad} \emph
  {et~al.}}]{ATLAS:2011qdp}%
  \BibitemOpen
  \bibfield  {author} {\bibinfo {author} {\bibfnamefont {G.}~\bibnamefont
  {Aad}} \emph {et~al.} (\bibinfo {collaboration} {ATLAS}),\ }\bibfield
  {title} {\bibinfo {title} {{Measurement of the inclusive $W^\pm$ and Z/gamma
  cross sections in the electron and muon decay channels in $pp$ collisions at
  $\sqrt{s}=7$ TeV with the ATLAS detector}},\ }\href
  {https://doi.org/10.1103/PhysRevD.85.072004} {\bibfield  {journal} {\bibinfo
  {journal} {Phys. Rev. D}\ }\textbf {\bibinfo {volume} {85}},\ \bibinfo
  {pages} {072004} (\bibinfo {year} {2012})},\ \Eprint
  {https://arxiv.org/abs/1109.5141} {arXiv:1109.5141 [hep-ex]} \BibitemShut
  {NoStop}%
\bibitem [{\citenamefont {Aaboud}\ \emph {et~al.}(2017)\citenamefont {Aaboud}
  \emph {et~al.}}]{ATLAS:2016nqi}%
  \BibitemOpen
  \bibfield  {author} {\bibinfo {author} {\bibfnamefont {M.}~\bibnamefont
  {Aaboud}} \emph {et~al.} (\bibinfo {collaboration} {ATLAS}),\ }\bibfield
  {title} {\bibinfo {title} {{Precision measurement and interpretation of
  inclusive $W^+$ , $W^-$ and $Z/\gamma ^*$ production cross sections with the
  ATLAS detector}},\ }\href {https://doi.org/10.1140/epjc/s10052-017-4911-9}
  {\bibfield  {journal} {\bibinfo  {journal} {Eur. Phys. J. C}\ }\textbf
  {\bibinfo {volume} {77}},\ \bibinfo {pages} {367} (\bibinfo {year} {2017})},\
  \Eprint {https://arxiv.org/abs/1612.03016} {arXiv:1612.03016 [hep-ex]}
  \BibitemShut {NoStop}%
\bibitem [{\citenamefont {Aad}\ \emph {et~al.}(2019{\natexlab{a}})\citenamefont
  {Aad} \emph {et~al.}}]{ATLAS:2019fgb}%
  \BibitemOpen
  \bibfield  {author} {\bibinfo {author} {\bibfnamefont {G.}~\bibnamefont
  {Aad}} \emph {et~al.} (\bibinfo {collaboration} {ATLAS}),\ }\bibfield
  {title} {\bibinfo {title} {{Measurement of the cross-section and charge
  asymmetry of $W$ bosons produced in proton\textendash{}proton collisions at
  $\sqrt{s}=8~\text {TeV}$ with the ATLAS detector}},\ }\href
  {https://doi.org/10.1140/epjc/s10052-019-7199-0} {\bibfield  {journal}
  {\bibinfo  {journal} {Eur. Phys. J. C}\ }\textbf {\bibinfo {volume} {79}},\
  \bibinfo {pages} {760} (\bibinfo {year} {2019}{\natexlab{a}})},\ \Eprint
  {https://arxiv.org/abs/1904.05631} {arXiv:1904.05631 [hep-ex]} \BibitemShut
  {NoStop}%
\bibitem [{\citenamefont {Chatrchyan}\ \emph {et~al.}(2011)\citenamefont
  {Chatrchyan} \emph {et~al.}}]{CMS:2011aa}%
  \BibitemOpen
  \bibfield  {author} {\bibinfo {author} {\bibfnamefont {S.}~\bibnamefont
  {Chatrchyan}} \emph {et~al.} (\bibinfo {collaboration} {CMS}),\ }\bibfield
  {title} {\bibinfo {title} {{Measurement of the Inclusive $W$ and $Z$
  Production Cross Sections in $pp$ Collisions at $\sqrt{s}=7$ TeV}},\ }\href
  {https://doi.org/10.1007/JHEP10(2011)132} {\bibfield  {journal} {\bibinfo
  {journal} {JHEP}\ }\textbf {\bibinfo {volume} {10}},\ \bibinfo {pages}
  {132}},\ \Eprint {https://arxiv.org/abs/1107.4789} {arXiv:1107.4789 [hep-ex]}
  \BibitemShut {NoStop}%
\bibitem [{\citenamefont {Chatrchyan}\ \emph
  {et~al.}(2012{\natexlab{a}})\citenamefont {Chatrchyan} \emph
  {et~al.}}]{CMS:2012ivw}%
  \BibitemOpen
  \bibfield  {author} {\bibinfo {author} {\bibfnamefont {S.}~\bibnamefont
  {Chatrchyan}} \emph {et~al.} (\bibinfo {collaboration} {CMS}),\ }\bibfield
  {title} {\bibinfo {title} {{Measurement of the Electron Charge Asymmetry in
  Inclusive $W$ Production in $pp$ Collisions at $\sqrt{s}=7$ TeV}},\ }\href
  {https://doi.org/10.1103/PhysRevLett.109.111806} {\bibfield  {journal}
  {\bibinfo  {journal} {Phys. Rev. Lett.}\ }\textbf {\bibinfo {volume} {109}},\
  \bibinfo {pages} {111806} (\bibinfo {year} {2012}{\natexlab{a}})},\ \Eprint
  {https://arxiv.org/abs/1206.2598} {arXiv:1206.2598 [hep-ex]} \BibitemShut
  {NoStop}%
\bibitem [{\citenamefont {Chatrchyan}\ \emph
  {et~al.}(2012{\natexlab{b}})\citenamefont {Chatrchyan} \emph
  {et~al.}}]{CMS:2012fgk}%
  \BibitemOpen
  \bibfield  {author} {\bibinfo {author} {\bibfnamefont {S.}~\bibnamefont
  {Chatrchyan}} \emph {et~al.} (\bibinfo {collaboration} {CMS}),\ }\bibfield
  {title} {\bibinfo {title} {{Study of $W$ boson production in PbPb and $pp$
  collisions at $\sqrt{s_{NN}}=2.76$ TeV}},\ }\href
  {https://doi.org/10.1016/j.physletb.2012.07.025} {\bibfield  {journal}
  {\bibinfo  {journal} {Phys. Lett. B}\ }\textbf {\bibinfo {volume} {715}},\
  \bibinfo {pages} {66} (\bibinfo {year} {2012}{\natexlab{b}})},\ \Eprint
  {https://arxiv.org/abs/1205.6334} {arXiv:1205.6334 [nucl-ex]} \BibitemShut
  {NoStop}%
\bibitem [{\citenamefont {Chatrchyan}\ \emph {et~al.}(2014)\citenamefont
  {Chatrchyan} \emph {et~al.}}]{CMS:2013pzl}%
  \BibitemOpen
  \bibfield  {author} {\bibinfo {author} {\bibfnamefont {S.}~\bibnamefont
  {Chatrchyan}} \emph {et~al.} (\bibinfo {collaboration} {CMS}),\ }\bibfield
  {title} {\bibinfo {title} {{Measurement of the Muon Charge Asymmetry in
  Inclusive $pp \to W+X$ Production at $\sqrt s =$ 7 TeV and an Improved
  Determination of Light Parton Distribution Functions}},\ }\href
  {https://doi.org/10.1103/PhysRevD.90.032004} {\bibfield  {journal} {\bibinfo
  {journal} {Phys. Rev. D}\ }\textbf {\bibinfo {volume} {90}},\ \bibinfo
  {pages} {032004} (\bibinfo {year} {2014})},\ \Eprint
  {https://arxiv.org/abs/1312.6283} {arXiv:1312.6283 [hep-ex]} \BibitemShut
  {NoStop}%
\bibitem [{\citenamefont {Khachatryan}\ \emph {et~al.}(2016)\citenamefont
  {Khachatryan} \emph {et~al.}}]{CMS:2016qqr}%
  \BibitemOpen
  \bibfield  {author} {\bibinfo {author} {\bibfnamefont {V.}~\bibnamefont
  {Khachatryan}} \emph {et~al.} (\bibinfo {collaboration} {CMS}),\ }\bibfield
  {title} {\bibinfo {title} {{Measurement of the differential cross section and
  charge asymmetry for inclusive $\mathrm {p}\mathrm {p}\rightarrow \mathrm
  {W}^{\pm }+X$ production at ${\sqrt{s}} = 8$ TeV}},\ }\href
  {https://doi.org/10.1140/epjc/s10052-016-4293-4} {\bibfield  {journal}
  {\bibinfo  {journal} {Eur. Phys. J. C}\ }\textbf {\bibinfo {volume} {76}},\
  \bibinfo {pages} {469} (\bibinfo {year} {2016})},\ \Eprint
  {https://arxiv.org/abs/1603.01803} {arXiv:1603.01803 [hep-ex]} \BibitemShut
  {NoStop}%
\bibitem [{\citenamefont {Aaij}\ \emph {et~al.}(2012)\citenamefont {Aaij} \emph
  {et~al.}}]{LHCb:2012lki}%
  \BibitemOpen
  \bibfield  {author} {\bibinfo {author} {\bibfnamefont {R.}~\bibnamefont
  {Aaij}} \emph {et~al.} (\bibinfo {collaboration} {LHCb}),\ }\bibfield
  {title} {\bibinfo {title} {{Inclusive $W$ and $Z$ production in the forward
  region at $\sqrt{s} = 7$ TeV}},\ }\href
  {https://doi.org/10.1007/JHEP06(2012)058} {\bibfield  {journal} {\bibinfo
  {journal} {JHEP}\ }\textbf {\bibinfo {volume} {06}},\ \bibinfo {pages}
  {058}},\ \Eprint {https://arxiv.org/abs/1204.1620} {arXiv:1204.1620 [hep-ex]}
  \BibitemShut {NoStop}%
\bibitem [{\citenamefont {Aaij}\ \emph {et~al.}(2014)\citenamefont {Aaij} \emph
  {et~al.}}]{LHCb:2014liz}%
  \BibitemOpen
  \bibfield  {author} {\bibinfo {author} {\bibfnamefont {R.}~\bibnamefont
  {Aaij}} \emph {et~al.} (\bibinfo {collaboration} {LHCb}),\ }\bibfield
  {title} {\bibinfo {title} {{Measurement of the forward $W$ boson
  cross-section in $pp$ collisions at $\sqrt{s} = 7 {\rm \, TeV}$}},\ }\href
  {https://doi.org/10.1007/JHEP12(2014)079} {\bibfield  {journal} {\bibinfo
  {journal} {JHEP}\ }\textbf {\bibinfo {volume} {12}},\ \bibinfo {pages}
  {079}},\ \Eprint {https://arxiv.org/abs/1408.4354} {arXiv:1408.4354 [hep-ex]}
  \BibitemShut {NoStop}%
\bibitem [{\citenamefont {Aaij}\ \emph
  {et~al.}(2016{\natexlab{a}})\citenamefont {Aaij} \emph
  {et~al.}}]{LHCb:2015mad}%
  \BibitemOpen
  \bibfield  {author} {\bibinfo {author} {\bibfnamefont {R.}~\bibnamefont
  {Aaij}} \emph {et~al.} (\bibinfo {collaboration} {LHCb}),\ }\bibfield
  {title} {\bibinfo {title} {{Measurement of forward W and Z boson production
  in $pp$ collisions at $ \sqrt{s}=8 $ TeV}},\ }\href
  {https://doi.org/10.1007/JHEP01(2016)155} {\bibfield  {journal} {\bibinfo
  {journal} {JHEP}\ }\textbf {\bibinfo {volume} {01}},\ \bibinfo {pages}
  {155}},\ \Eprint {https://arxiv.org/abs/1511.08039} {arXiv:1511.08039
  [hep-ex]} \BibitemShut {NoStop}%
\bibitem [{\citenamefont {Aaij}\ \emph
  {et~al.}(2016{\natexlab{b}})\citenamefont {Aaij} \emph
  {et~al.}}]{LHCb:2016zpq}%
  \BibitemOpen
  \bibfield  {author} {\bibinfo {author} {\bibfnamefont {R.}~\bibnamefont
  {Aaij}} \emph {et~al.} (\bibinfo {collaboration} {LHCb}),\ }\bibfield
  {title} {\bibinfo {title} {{Measurement of forward $W\to e\nu$ production in
  $pp$ collisions at $\sqrt{s}=8\,$TeV}},\ }\href
  {https://doi.org/10.1007/JHEP10(2016)030} {\bibfield  {journal} {\bibinfo
  {journal} {JHEP}\ }\textbf {\bibinfo {volume} {10}},\ \bibinfo {pages}
  {030}},\ \Eprint {https://arxiv.org/abs/1608.01484} {arXiv:1608.01484
  [hep-ex]} \BibitemShut {NoStop}%
\bibitem [{\citenamefont {Aad}\ \emph {et~al.}(2019{\natexlab{b}})\citenamefont
  {Aad} \emph {et~al.}}]{ATLAS:2019fyu}%
  \BibitemOpen
  \bibfield  {author} {\bibinfo {author} {\bibfnamefont {G.}~\bibnamefont
  {Aad}} \emph {et~al.} (\bibinfo {collaboration} {ATLAS}),\ }\bibfield
  {title} {\bibinfo {title} {{Measurement of $W^{\pm }$-boson and Z-boson
  production cross-sections in pp collisions at $\sqrt{s}=2.76$ TeV with the
  ATLAS detector}},\ }\href {https://doi.org/10.1140/epjc/s10052-019-7399-7}
  {\bibfield  {journal} {\bibinfo  {journal} {Eur. Phys. J. C}\ }\textbf
  {\bibinfo {volume} {79}},\ \bibinfo {pages} {901} (\bibinfo {year}
  {2019}{\natexlab{b}})},\ \Eprint {https://arxiv.org/abs/1907.03567}
  {arXiv:1907.03567 [hep-ex]} \BibitemShut {NoStop}%
\bibitem [{\citenamefont {Aaboud}\ \emph {et~al.}(2019)\citenamefont {Aaboud}
  \emph {et~al.}}]{ATLAS:2018pyl}%
  \BibitemOpen
  \bibfield  {author} {\bibinfo {author} {\bibfnamefont {M.}~\bibnamefont
  {Aaboud}} \emph {et~al.} (\bibinfo {collaboration} {ATLAS}),\ }\bibfield
  {title} {\bibinfo {title} {{Measurements of $W$ and $Z$ boson production in
  $pp$ collisions at $\sqrt{s}=5.02$ TeV with the ATLAS detector}},\ }\href
  {https://doi.org/10.1140/epjc/s10052-019-6622-x} {\bibfield  {journal}
  {\bibinfo  {journal} {Eur. Phys. J. C}\ }\textbf {\bibinfo {volume} {79}},\
  \bibinfo {pages} {128} (\bibinfo {year} {2019})},\ \bibinfo {note} {[Erratum:
  Eur.Phys.J.C 79, 374 (2019)]},\ \Eprint {https://arxiv.org/abs/1810.08424}
  {arXiv:1810.08424 [hep-ex]} \BibitemShut {NoStop}%
\bibitem [{\citenamefont {Adam}\ \emph {et~al.}(2017)\citenamefont {Adam} \emph
  {et~al.}}]{ALICE:2016rzo}%
  \BibitemOpen
  \bibfield  {author} {\bibinfo {author} {\bibfnamefont {J.}~\bibnamefont
  {Adam}} \emph {et~al.} (\bibinfo {collaboration} {ALICE}),\ }\bibfield
  {title} {\bibinfo {title} {{W and Z boson production in p-Pb collisions at
  $\sqrt{s_{\rm NN}}$ = 5.02 TeV}},\ }\href
  {https://doi.org/10.1007/JHEP02(2017)077} {\bibfield  {journal} {\bibinfo
  {journal} {JHEP}\ }\textbf {\bibinfo {volume} {02}},\ \bibinfo {pages}
  {077}},\ \Eprint {https://arxiv.org/abs/1611.03002} {arXiv:1611.03002
  [nucl-ex]} \BibitemShut {NoStop}%
\bibitem [{\citenamefont {{ALICE collaboration}}()}]{ALICE:2022cxs}%
  \BibitemOpen
  \bibfield  {author} {\bibinfo {author} {\bibnamefont {{ALICE
  collaboration}}},\ }\href@noop {} {\bibinfo {title} {{W$^\pm$-boson
  production in p$-$Pb collisions at $\sqrt{s_{NN}} = 8.16$ TeV and PbPb
  collisions at $\sqrt{s_{NN}} = 5.02$ TeV}}},\ \Eprint
  {https://arxiv.org/abs/2204.10640} {arXiv:2204.10640 [nucl-ex]} \BibitemShut
  {NoStop}%
\bibitem [{\citenamefont {Sirunyan}\ \emph {et~al.}(2020)\citenamefont
  {Sirunyan} \emph {et~al.}}]{CMS:2019leu}%
  \BibitemOpen
  \bibfield  {author} {\bibinfo {author} {\bibfnamefont {A.~M.}\ \bibnamefont
  {Sirunyan}} \emph {et~al.} (\bibinfo {collaboration} {CMS}),\ }\bibfield
  {title} {\bibinfo {title} {{Observation of nuclear modifications in W$^\pm$
  boson production in pPb collisions at $\sqrt{s_\mathrm{NN}} =$ 8.16 TeV}},\
  }\href {https://doi.org/10.1016/j.physletb.2019.135048} {\bibfield  {journal}
  {\bibinfo  {journal} {Phys. Lett. B}\ }\textbf {\bibinfo {volume} {800}},\
  \bibinfo {pages} {135048} (\bibinfo {year} {2020})},\ \Eprint
  {https://arxiv.org/abs/1905.01486} {arXiv:1905.01486 [hep-ex]} \BibitemShut
  {NoStop}%
\bibitem [{\citenamefont {Aad}\ \emph {et~al.}(2015)\citenamefont {Aad} \emph
  {et~al.}}]{ATLAS:2014sic}%
  \BibitemOpen
  \bibfield  {author} {\bibinfo {author} {\bibfnamefont {G.}~\bibnamefont
  {Aad}} \emph {et~al.} (\bibinfo {collaboration} {ATLAS}),\ }\bibfield
  {title} {\bibinfo {title} {{Measurement of the production and lepton charge
  asymmetry of $W$ bosons in Pb+Pb collisions at $\mathbf {\sqrt{\mathbf
  {s}_{\mathrm {\mathbf {NN}}}}=2.76\;TeV}$ with the ATLAS detector}},\ }\href
  {https://doi.org/10.1140/epjc/s10052-014-3231-6} {\bibfield  {journal}
  {\bibinfo  {journal} {Eur. Phys. J. C}\ }\textbf {\bibinfo {volume} {75}},\
  \bibinfo {pages} {23} (\bibinfo {year} {2015})},\ \Eprint
  {https://arxiv.org/abs/1408.4674} {arXiv:1408.4674 [hep-ex]} \BibitemShut
  {NoStop}%
\bibitem [{\citenamefont {Aad}\ \emph {et~al.}(2019{\natexlab{c}})\citenamefont
  {Aad} \emph {et~al.}}]{ATLAS:2019ibd}%
  \BibitemOpen
  \bibfield  {author} {\bibinfo {author} {\bibfnamefont {G.}~\bibnamefont
  {Aad}} \emph {et~al.} (\bibinfo {collaboration} {ATLAS}),\ }\bibfield
  {title} {\bibinfo {title} {{Measurement of $W^\pm $ boson production in Pb+Pb
  collisions at $\sqrt{s_{\mathrm{NN}}} = 5.02~\text {Te}\text {V}$ with the
  ATLAS detector}},\ }\href {https://doi.org/10.1140/epjc/s10052-019-7439-3}
  {\bibfield  {journal} {\bibinfo  {journal} {Eur. Phys. J. C}\ }\textbf
  {\bibinfo {volume} {79}},\ \bibinfo {pages} {935} (\bibinfo {year}
  {2019}{\natexlab{c}})},\ \Eprint {https://arxiv.org/abs/1907.10414}
  {arXiv:1907.10414 [nucl-ex]} \BibitemShut {NoStop}%
\bibitem [{\citenamefont {Dulat}\ \emph {et~al.}(2016)\citenamefont {Dulat},
  \citenamefont {Hou}, \citenamefont {Gao}, \citenamefont {Guzzi},
  \citenamefont {Huston}, \citenamefont {Nadolsky}, \citenamefont {Pumplin},
  \citenamefont {Schmidt}, \citenamefont {Stump},\ and\ \citenamefont
  {Yuan}}]{Dulat:2015mca}%
  \BibitemOpen
  \bibfield  {author} {\bibinfo {author} {\bibfnamefont {S.}~\bibnamefont
  {Dulat}}, \bibinfo {author} {\bibfnamefont {T.-J.}\ \bibnamefont {Hou}},
  \bibinfo {author} {\bibfnamefont {J.}~\bibnamefont {Gao}}, \bibinfo {author}
  {\bibfnamefont {M.}~\bibnamefont {Guzzi}}, \bibinfo {author} {\bibfnamefont
  {J.}~\bibnamefont {Huston}}, \bibinfo {author} {\bibfnamefont
  {P.}~\bibnamefont {Nadolsky}}, \bibinfo {author} {\bibfnamefont
  {J.}~\bibnamefont {Pumplin}}, \bibinfo {author} {\bibfnamefont
  {C.}~\bibnamefont {Schmidt}}, \bibinfo {author} {\bibfnamefont
  {D.}~\bibnamefont {Stump}},\ and\ \bibinfo {author} {\bibfnamefont {C.~P.}\
  \bibnamefont {Yuan}},\ }\bibfield  {title} {\bibinfo {title} {{New parton
  distribution functions from a global analysis of quantum chromodynamics}},\
  }\href {https://doi.org/10.1103/PhysRevD.93.033006} {\bibfield  {journal}
  {\bibinfo  {journal} {Phys. Rev. D}\ }\textbf {\bibinfo {volume} {93}},\
  \bibinfo {pages} {033006} (\bibinfo {year} {2016})},\ \Eprint
  {https://arxiv.org/abs/1506.07443} {arXiv:1506.07443 [hep-ph]} \BibitemShut
  {NoStop}%
\bibitem [{\citenamefont {Eskola}\ \emph {et~al.}(2009)\citenamefont {Eskola},
  \citenamefont {Paukkunen},\ and\ \citenamefont {Salgado}}]{Eskola:2009uj}%
  \BibitemOpen
  \bibfield  {author} {\bibinfo {author} {\bibfnamefont {K.~J.}\ \bibnamefont
  {Eskola}}, \bibinfo {author} {\bibfnamefont {H.}~\bibnamefont {Paukkunen}},\
  and\ \bibinfo {author} {\bibfnamefont {C.~A.}\ \bibnamefont {Salgado}},\
  }\bibfield  {title} {\bibinfo {title} {{EPS09: A New Generation of NLO and LO
  Nuclear Parton Distribution Functions}},\ }\href
  {https://doi.org/10.1088/1126-6708/2009/04/065} {\bibfield  {journal}
  {\bibinfo  {journal} {JHEP}\ }\textbf {\bibinfo {volume} {04}},\ \bibinfo
  {pages} {065}},\ \Eprint {https://arxiv.org/abs/0902.4154} {arXiv:0902.4154
  [hep-ph]} \BibitemShut {NoStop}%
\bibitem [{\citenamefont {Kusina}\ \emph {et~al.}(2017)\citenamefont {Kusina},
  \citenamefont {Lyonnet}, \citenamefont {Clark}, \citenamefont {Godat},
  \citenamefont {Jezo}, \citenamefont {Kovarik}, \citenamefont {Olness},
  \citenamefont {Schienbein},\ and\ \citenamefont {Yu}}]{Kusina:2016fxy}%
  \BibitemOpen
  \bibfield  {author} {\bibinfo {author} {\bibfnamefont {A.}~\bibnamefont
  {Kusina}}, \bibinfo {author} {\bibfnamefont {F.}~\bibnamefont {Lyonnet}},
  \bibinfo {author} {\bibfnamefont {D.~B.}\ \bibnamefont {Clark}}, \bibinfo
  {author} {\bibfnamefont {E.}~\bibnamefont {Godat}}, \bibinfo {author}
  {\bibfnamefont {T.}~\bibnamefont {Jezo}}, \bibinfo {author} {\bibfnamefont
  {K.}~\bibnamefont {Kovarik}}, \bibinfo {author} {\bibfnamefont {F.~I.}\
  \bibnamefont {Olness}}, \bibinfo {author} {\bibfnamefont {I.}~\bibnamefont
  {Schienbein}},\ and\ \bibinfo {author} {\bibfnamefont {J.~Y.}\ \bibnamefont
  {Yu}},\ }\bibfield  {title} {\bibinfo {title} {{Vector boson production in
  pPb and PbPb collisions at the LHC and its impact on nCTEQ15 PDFs}},\ }\href
  {https://doi.org/10.1140/epjc/s10052-017-5036-x} {\bibfield  {journal}
  {\bibinfo  {journal} {Eur. Phys. J. C}\ }\textbf {\bibinfo {volume} {77}},\
  \bibinfo {pages} {488} (\bibinfo {year} {2017})},\ \Eprint
  {https://arxiv.org/abs/1610.02925} {arXiv:1610.02925 [nucl-th]} \BibitemShut
  {NoStop}%
\bibitem [{\citenamefont {Eskola}\ \emph {et~al.}(2017)\citenamefont {Eskola},
  \citenamefont {Paakkinen}, \citenamefont {Paukkunen},\ and\ \citenamefont
  {Salgado}}]{Eskola:2016oht}%
  \BibitemOpen
  \bibfield  {author} {\bibinfo {author} {\bibfnamefont {K.~J.}\ \bibnamefont
  {Eskola}}, \bibinfo {author} {\bibfnamefont {P.}~\bibnamefont {Paakkinen}},
  \bibinfo {author} {\bibfnamefont {H.}~\bibnamefont {Paukkunen}},\ and\
  \bibinfo {author} {\bibfnamefont {C.~A.}\ \bibnamefont {Salgado}},\
  }\bibfield  {title} {\bibinfo {title} {{EPPS16: Nuclear parton distributions
  with LHC data}},\ }\href {https://doi.org/10.1140/epjc/s10052-017-4725-9}
  {\bibfield  {journal} {\bibinfo  {journal} {Eur. Phys. J. C}\ }\textbf
  {\bibinfo {volume} {77}},\ \bibinfo {pages} {163} (\bibinfo {year} {2017})},\
  \Eprint {https://arxiv.org/abs/1612.05741} {arXiv:1612.05741 [hep-ph]}
  \BibitemShut {NoStop}%
\bibitem [{\citenamefont {Sa}\ \emph {et~al.}(2012)\citenamefont {Sa},
  \citenamefont {Zhou}, \citenamefont {Yan}, \citenamefont {Li}, \citenamefont
  {Feng}, \citenamefont {Dong},\ and\ \citenamefont {Cai}}]{Sa:2011ye}%
  \BibitemOpen
  \bibfield  {author} {\bibinfo {author} {\bibfnamefont {B.-H.}\ \bibnamefont
  {Sa}}, \bibinfo {author} {\bibfnamefont {D.-M.}\ \bibnamefont {Zhou}},
  \bibinfo {author} {\bibfnamefont {Y.-L.}\ \bibnamefont {Yan}}, \bibinfo
  {author} {\bibfnamefont {X.-M.}\ \bibnamefont {Li}}, \bibinfo {author}
  {\bibfnamefont {S.-Q.}\ \bibnamefont {Feng}}, \bibinfo {author}
  {\bibfnamefont {B.-G.}\ \bibnamefont {Dong}},\ and\ \bibinfo {author}
  {\bibfnamefont {X.}~\bibnamefont {Cai}},\ }\bibfield  {title} {\bibinfo
  {title} {{PACIAE 2.0: An Updated parton and hadron cascade model (program)
  for the relativistic nuclear collisions}},\ }\href
  {https://doi.org/10.1016/j.cpc.2011.08.021} {\bibfield  {journal} {\bibinfo
  {journal} {Comput. Phys. Commun.}\ }\textbf {\bibinfo {volume} {183}},\
  \bibinfo {pages} {333} (\bibinfo {year} {2012})},\ \Eprint
  {https://arxiv.org/abs/1104.1238} {arXiv:1104.1238 [nucl-th]} \BibitemShut
  {NoStop}%
\bibitem [{\citenamefont {Sjostrand}\ \emph {et~al.}(2006)\citenamefont
  {Sjostrand}, \citenamefont {Mrenna},\ and\ \citenamefont
  {Skands}}]{Sjostrand:2006za}%
  \BibitemOpen
  \bibfield  {author} {\bibinfo {author} {\bibfnamefont {T.}~\bibnamefont
  {Sjostrand}}, \bibinfo {author} {\bibfnamefont {S.}~\bibnamefont {Mrenna}},\
  and\ \bibinfo {author} {\bibfnamefont {P.~Z.}\ \bibnamefont {Skands}},\
  }\bibfield  {title} {\bibinfo {title} {{PYTHIA 6.4 Physics and Manual}},\
  }\href {https://doi.org/10.1088/1126-6708/2006/05/026} {\bibfield  {journal}
  {\bibinfo  {journal} {JHEP}\ }\textbf {\bibinfo {volume} {05}},\ \bibinfo
  {pages} {026}},\ \Eprint {https://arxiv.org/abs/hep-ph/0603175}
  {arXiv:hep-ph/0603175} \BibitemShut {NoStop}%
\bibitem [{\citenamefont {Combridge}\ \emph {et~al.}(1977)\citenamefont
  {Combridge}, \citenamefont {Kripfganz},\ and\ \citenamefont
  {Ranft}}]{Combridge:1977dm}%
  \BibitemOpen
  \bibfield  {author} {\bibinfo {author} {\bibfnamefont {B.~L.}\ \bibnamefont
  {Combridge}}, \bibinfo {author} {\bibfnamefont {J.}~\bibnamefont
  {Kripfganz}},\ and\ \bibinfo {author} {\bibfnamefont {J.}~\bibnamefont
  {Ranft}},\ }\bibfield  {title} {\bibinfo {title} {{Hadron Production at Large
  Transverse Momentum and QCD}},\ }\href
  {https://doi.org/10.1016/0370-2693(77)90528-7} {\bibfield  {journal}
  {\bibinfo  {journal} {Phys. Lett. B}\ }\textbf {\bibinfo {volume} {70}},\
  \bibinfo {pages} {234} (\bibinfo {year} {1977})}\BibitemShut {NoStop}%
\bibitem [{\citenamefont {Field}(1989)}]{field}%
  \BibitemOpen
  \bibfield  {author} {\bibinfo {author} {\bibfnamefont {R.~D.}\ \bibnamefont
  {Field}},\ }\href@noop {} {\emph {\bibinfo {title} {Application of
  perturbative QCD}}}\ (\bibinfo  {publisher} {Addison-Wesley Publishing
  Company, Inc.},\ \bibinfo {year} {1989})\BibitemShut {NoStop}%
\bibitem [{Note1()}]{Note1}%
  \BibitemOpen
  \bibinfo {note} {The ALICE muon spectrometer covers negative pseudorapidity.
  However, in symmetric Pb+Pb collisions, positive values of rapidity are
  conventionally used for the muon coverage.}\BibitemShut {Stop}%
\bibitem [{\citenamefont {Conesa Del~Valle}(2007)}]{ConesaDelValle:2007dza}%
  \BibitemOpen
  \bibfield  {author} {\bibinfo {author} {\bibfnamefont {Z.}~\bibnamefont
  {Conesa Del~Valle}},\ }\emph {\bibinfo {title} {{Performance of the ALICE
  muon spectrometer. Weak boson production and measurement in heavy-ion
  collisions at LHC}}},\ \href@noop {} {Ph.D. thesis},\ \bibinfo  {school}
  {Nantes U.} (\bibinfo {year} {2007})\BibitemShut {NoStop}%
\end{thebibliography}
%

\end{document}